\documentclass[iop]{emulateapj}
\def\lsim{\lower.5ex\hbox{$\; \buildrel < \over \sim \;$}}
\def\gsim{\lower.5ex\hbox{$\; \buildrel > \over \sim \;$}}

\usepackage{natbib}
\slugcomment{}

\begin{document}

\title{How unusual is XRF 060218 ? }

\author { Samir Mandal and David Eichler}
\affil{Department of Physics, Ben-Gurion University of the Negev,
Beer-Sheva 84105, Israel; eichler@bgu.ac.il, mandals@bgu.ac.il}

\begin{abstract}
Light curves are calculated for an off-axis observer due the
scattering of primary radiation by extended baryonic material. The
unusually long duration and the chromaticity of the light curves above several KeV of
XRF 060218 can be explained as a result of the acceleration of the
baryonic scattering material by the primary radiation.  The observed
light curves by our model and detailed fits to the data are
presented. The model predicts that  $\sim 4\times 10^{48}$ ergs are put
into accelerated, mildly relativistic baryons by the
radiation pressure at large radii from the central engine. It is suggested that the emission below 3 KeV, which lies {\it below} the Amati relation, is a baryon contaminated fireball.
\end{abstract}

\section{Introduction}

Nearly all known GRB with known red shifts lie either on or to one side of the Amati relation, which is usually shown as a plot of GRB spectral peak energy $E_{peak}$, on the y-axis, as a function of the isotropic equivalent energy $E_{iso}$. This can be understood as being due to the observer's light of sight being  offset relative to the direction of motion of the material at which the emission or last scattering took place (\citealt{EL04, Yamazaki04}). However, the absence of GRB below the ``edge" - defined by the Amati relation to be $E_{peak} \sim [E_{iso}/10^{54}erg]^{1/2}$ MeV  - is curious. For it states that the isotropic equivalent photon number in a GRB decreases in proportion to the peak photon energy. It thus strongly constrains scatter in the extent of any process within a GRB fireball that reduces photon energy without decreasing the number of photons, e.g. adiabatic losses of photons emitted in an optically thick region. Similarly, it constrains scatter in the number of photons emitted in GBR having a given spectral peak, e.g. the number of energetic particles emitting in an optically thin region in a given magnetic field.

That the lower right side of the Amati graph is free of any GRB of known redshift is particularly remarkable considering that it represents the brightest GRBs at a given spectral peak. It suggests that there is no such thing as a ``slightly compromised" GRB whose photons are softened without being lost, e.g. as in a dirty fireball. In this letter, however, we suggest that the X-ray flash (XRF) 060218 is a dirty fireball. We note that, unlike most XRF, the low energy component has, despite its  much softer spectrum, a similar photon number to classical GRB, as opposed to most XRF, which have far fewer photons.

XRF 060218 was detected with the Burst Alert Telescope (BAT)
instrument onboard Swift spacecraft. The
spectrum peaks below 10 keV, thus classifying this transient as an XRF.
XRF 060218 is distinguished by its unprecedentedly long duration ($\sim 2000$ sec) with a
 smooth light curve.
The light curves show a significant spectral lag, with
 soft photons lagging behind the hard photons, as usually seen in long
 GRBs (\citealt{Norris00}). The peak of the light curves are at $405\pm25$,
 $735\pm9$, $919\pm7$ and $1082\pm13$ seconds (\citealt{Liang06})
 in the energy band (15-150) keV, (5-10) keV, (2-5) keV and (0.3-2) keV respectively.
 The XRT spectrum (0.3-10 keV) shows a thermal component in soft X-ray with temperature
 $KT\approx 0.17$ keV (\citealt{Campana06}). The high energy spectra
 (15-150 keV) from BAT show spectral softening with time.

 The optically discovered supernova 2006aj is
 associated with XRF 060218.
 The optical afterglow spectra and the strong emission
 lines from the host galaxy represent
 a redshift of $z=0.03342\pm0.00002$, corresponding to
 a distance of
 $\sim 140$ Mpc (\citealt{Pian06}).
 The isotropic equivalent prompt energy release is
 $E_{iso}=(6.2\pm0.3)\times 10^{49}$ erg (\citealt{Campana06}).

The unusually long duration of this event  raises the question of whether the central engine
goes on this long, or whether the duration represents something
downstream of the central engine,  e.g. shock breakout through
the surface of the star or a light echo from slowly moving, extended
material. The chromaticity of the duration of the event and the
location of the peak is significant. It challenges breakout and light echo models, in which the
duration is established by hydrodynamics. We focus on this issue.
We propose that the radiation
pressure of the photons on the matter can accelerate it up to
relativistic Lorentz factors, which makes the duration appear longer
than the actual activity of the central engine. We  show that the intrinsic duration of
central engine activity need not be pathologically long for an XRF, but appears
longer because of the relativistic motion of the scatterer.  We also show that the
apparent duration is expected to be  wavelength dependent, and we
fit the observed light curves.

\section{The Model}
We assume that the primary radiation is scattered by an extended
baryonic cloud of optical depth unity and that the scattered
radiation is seen by an observer at an angle $\theta$ with the motion of the scatterer (Fig. 1).
 The entire scattering cloud is within the cone of the primary hard radiation and the  observer
sees the hard radiation due to scattering from the cloud whereas the extended soft component
may also reach the observer  directly (see below).

 The scatterer itself is offset
 from the axis of the primary radiation  and scatters
 only the soft fringes of
 the primary photon jet. The off-axis
 observer  sees a fast rise, slow decay light curve
  due to acceleration of the scattering
 baryons by the primary
 radiation pressure which causes the beam of the scattered
 radiation to narrow, intensify (the rise), and
 finally narrow to
 below the offset angle (the decay).
 We have calculated the light curve for different
 energy band following \citealt{EM07}.
 It produces an energy dependent delay in the light
 curve because, for the same primary
 photon energy, the observer  sees a photon energy that
 decreases with time
 due to the acceleration of the scatterer (\citealt{EM08}).
 In reality the scatterer can be an extended one.
 This may prolong the light curve
 at all photon energies
(i.e. create a light echo)
 as the contribution from the closer  part of the cloud is less delayed
  than that from the furthest part of the cloud.

 We have calculated the light curves by
 assuming the primary photon spectrum to be a
 broken power-law  i.e.,
\begin{eqnarray}
N(E) & = &
\left \{
\begin{array}{ll}
E^{\alpha},  & \quad  E < \xi \\
E^{\beta},  & \quad E >  \xi,\\
\end{array}     \right.
\end{eqnarray}
where E is photon energy. We have taken the
low energy photon index
$\alpha=-0.9$ with a break at $\xi=9$ keV and the high
energy index
$\beta=-2.2$  to
explain the light curves of XRF 060218. This is much softer than
assumed in previous papers, and the assumption is that the {\it
primary} radiation is that of an X-ray flash. The picture is that the scattering material itself is off the GRB jet axis, and sees an X-ray flash, which it then further scatters into the line of sight of the observer.
The duration of the primary radiation, as
seen by the scatterer, may then be itself prolonged by kinematic
effects relative to  the actual duration of activity of the central
engine. Moreover, the primary emission may contain in part
an intrinsically longer event such as shock breakout from the
surface of the host star, or a  ``dirty fireball'', either of which would be emitted over a larger angle than that of the  a baryon-poor  inner jet.

The primary source isotropic luminosity is taken to be  $1.2\times 10^{47}$
erg/s (\citealt{Liang06}). The luminosity of the source starts to decay exponentially
after 400 second with a decay constant 300 second.
The overall duration of the source activity is
590 sec which corresponds to 2500 sec in the observer frame.
We assume that the baryonic cloud is moving radially within the primary radiation cone
with an initial radial expansion speed $\beta=0.5$ and the centre of the cloud makes an
angle $\theta=20^\circ$ with respect to the observer.
We have considered an extended scattering cloud of diameter 400 light second
which, at an initial distance of $2\times 10^{13}$ cm
from the source, makes an opening angle $16.6^\circ$, so that  
the closest point of the scatterer moves in a direction $3.4^\circ$  off the line of sight.
All the primary radiation that hits the cloud is assumed to have been scattered
isotropically in the frame of the cloud by the time it exits.
In Fig. 2 we plot the observational data (histograms) with the model
light curves (solid lines) from the scattered primary emission, illustrated by the blue cone in figure 1.
The red, green, blue, magenta color
represent the data in the energy band (0.3-2) keV, (2-5) keV, (5-10)
keV and (15-150) keV respectively. The model light curves show a
good agreement with the observation except the very soft component
(red line) which is at least one order of magnitude smaller than the
observed value. So one needs an extra
source of soft component radiation to explain the observation, 
and this is illustrated by the wide red cone in figure 1.

In subsequent figures, we explore the dependence of different
parameters of the model on the resulting light curve. We compare the
light curves in two different energy bands (2-5 keV) and (15-150
keV) by changing the model parameters. In Fig. (3-5), we use red and
magenta lines to represent the light curves in (2-5 keV) and
(15-150 keV) energy band respectively with the same set of
parameters as in Fig. 2. Also the green and blue lines in Fig. (3-5)
represent the the light curves in (2-5 keV) and (15-150 keV)
energy band respectively but with a change of one parameter from
Fig. 2.  Fig. 3 shows the light curves for observing angle
$\theta=20^\circ$ (red and magenta) and $\theta=25^\circ$ (green and
blue). Clearly, as the observing angle increases the pulses rise,
peak,  and decay faster, and the observed flux decreases.

The light curve profile is sensitive to nature of the assumed
primary spectrum. The relative contributions between the pulses of
different energy bands are sensitive to both the low energy index
($\alpha$) and high energy power law index ($\beta$) of the primary
spectrum. Fig. 4 shows that as the low energy index decreases from
$\alpha=-0.9$ (red and magenta) to $\alpha=-1.1$ (green and blue),
the low energy flux decreases and the pulse peak shifts to an earlier time
and decays more rapidly. This change in low energy index will decrease the
normalization at the break energy ($\xi$=9 keV) and so the flux of
the high energy pulse also decreases. Fig. 5 shows the effect of high
energy index $\beta$ on the light curve as it increases from $\beta =-2.2$
(red and magenta) to $\beta=-2.0$ (green and blue). The high energy index
$\beta$ has almost the same effect as $\alpha$ on the light curve in
high energy band but the low energy light curve is not sensitive to
$\beta$.  As before, the low energy pulses are
unaffected by changing $\xi$ to high energy since the break
energy is well above the
low energy band. But the high energy contribution decreases
as $\xi$ increases since it reduces the high energy band.

The overall amount of kinetic energy imparted to the cloud  for the above choice of parameters is $3.8 \times 10^{48}$ ergs. This is in good agreement with the value estimated from radio afterglow calorimetry (\citealt{Soderberg04}), provided that there is no other component of blast energy. Perhaps most baryon kinetic energy, then, is from entrained baryons that have been accelerated by an otherwise baryon poor jet. There could be significant scatter in the extent of such entrainment, and the small value for kinetic energy obtained from this very close XRF, relative to cosmologically distant GRB, may simply be due to selection of brighter afterglows for  the more distant events. The same selection may of course apply to prompt emission, as nearby XRF and GRB are generally underluminous relative to distant ones.

\section{Concluding remarks}
We have accounted for both the {\it chromaticity}
of the light curve of GRB 060218/SN2006aj component
and the exceptionally long duration (relative to most
GRB and X-ray flashes), as being a result of our seeing
the photons after they have scattered off baryonic material
(e.g. wind material) that is
 accelerated by the radiation pressure of primary protons.
 The scatterer may lie
 at the periphery of the the primary jet of photons.
 These assumptions are quite reasonable:
 one would expect material blown out of the path of the
 primary GRB emission to be found at the periphery of the
 latter. An optical depth of order unity at a
 distance of order $10^{13.5}$cm, which we have tacitly
 assumed in order to match
 the amount of scattered radiation with the observed fluence,
 is somewhat high for a
 typical Wolf-Rayet wind, but, just prior to a supernova, the optical
 depth of the wind
 could be atypically high.
 Moreover, if   the material includes matter from the body of the progenitor itself that
 is ejected in the early stages of, or just prior to the GRB/SN, then large transverse
 gradients are expected near the periphery of the jet, and one expects some region to
 be of optical depth of order unity.  Even if the optical depth is greater than unity, the
 photons can nevertheless escape by reflecting into the backward hemisphere.
 Once the matter is accelerated to relativistic velocities, it is mostly  beamed forward
 even if it reflects back into the backward hemisphere in the frame of the scattering
 material.  While there are many parameters in the model, it should be recalled that
 there are several  nearby GRB-Sn associations, each with different high energy
 emission properties, and we suggest that the differences can be accounted for by
 variations in the parameters of the scattering material.

Because we have used  a very soft X-ray spectrum, we conclude that
our model may be spectrally consistent with a breakout shock
from the supernova together with a GRB. However, the total isotropic
equivalent energy, $\sim 6 \times 10^{49}$ ergs, is probably too
high for a breakout shock from the envelope, and may instead be powered by a GRB within.

The preferred alternative to a breakout shock is that the soft component is  the emission of a dirty fireball, as it contains the isotropic equivalent of nearly $10^{59}$ photons, close to that of a typical GRB, which peaks at a much higher energy. The KeV photons that dominate the total emitted energy can be interpreted as GRB photons that have been adiabatically decelerated while being trapped in the baryons.

That the peak energy is so far below that of a typical GRB is curious. It still leaves open the question of why there are still no examples of GRB in the lower right of the Amati graph that are closer to GRB in $(E_{peak}, E_{iso}$) space. Perhaps there is a sharp distinction between the baryon rich outflow, which traps and adiabatically decelerates photons, and the baryon poor outflow, which does not. Adiabatic deceleration would then take place either to a great extent, or to a very small extent.

The model and fit illustrate that significant
quantities of matter may be injected into a fireball at large
distance from the central engine and accelerated by the fireball to
relativistic energies. That the energetics are modest here may be a
result of the source being very close and viewed from a large
off-axis angle, and that even the scattering material may have been
somewhat off-axis. This is consistent with the otherwise
coincidental situation that GRB 060218 is both one of the longest,
closest, and softest gamma ray bursts. Thus, the various
distinguishing features of this GRB can be attributed to a large
off-axis viewing angle.

\acknowledgments
We thank C. Wheeler, E. Ramirez-Ruiz, A. Soderberg, A. McFadden,  and D. Frail for helpful discussions.
This research was supported by the Israel-US Binational Science
Foundation, the Israeli Academy of Science, and The Joan and Robert
Arnow Chair of Theoretical Astrophysics, and, via the Kavli
institute of Theoretical Physics, the US NSF grant NSF PHYS05-51164.

{}

\clearpage
\vfil\eject
\begin{figure}
\plotone{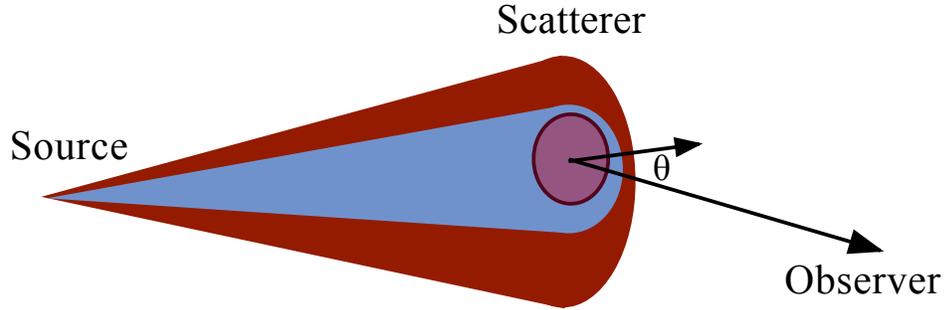}
\caption{The geometry of the model is shown.
The scattering material is colored purple and is
located near the periphery of the primary radiation cone,
where the primary spectrum appears soft due to kinematical
effects, as in the off-axis viewing model for X-ray flashes.
It is accelerated in the radial direction by the primary
radiation. Observer sees the hard radiation (blue narrow cone) due to scattering
from the cloud whereas  the extended soft component (red outside cone) is coming directly.
 In the text, the  ``offset angle" $\theta$ is relative to the axis of the scattering material.}
\end{figure}

\vfil\eject
\begin{figure}
\plotone{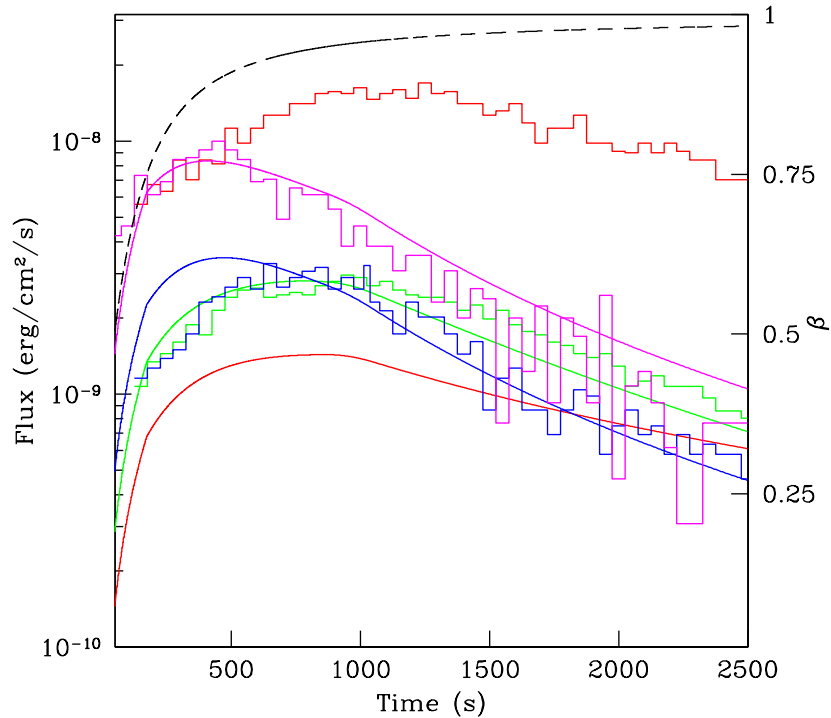}
\caption{Fitting of XRF 060218 light curves (histograms) with off-axis scattering model
(solid lines). The red, green, blue, magenta color represent the data in the energy
band (0.3-2) keV, (2-5) keV, (5-10) keV and (15-150) keV respectively.
For parameters see text. The observational data has been taken from \citealt{Liang06}.
The velocity $\beta$ (vertical axis right to the graph) of the scatterer is plotted by dashed line (black color). This indicates that the long duration is due to the acceleration of
the scatterer by the primary radiation.}
\end{figure}

\vfil\eject
\begin{figure}
\plotone{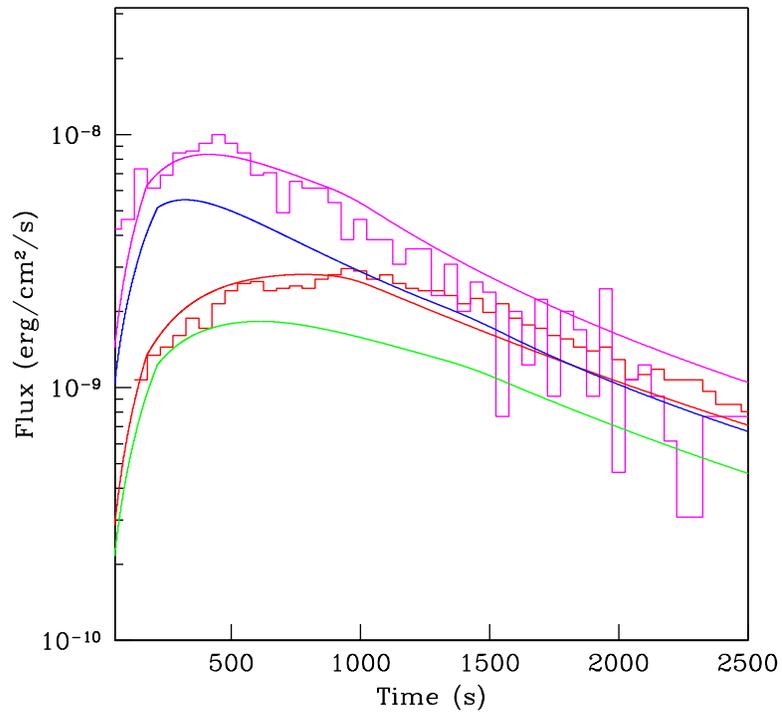}
\caption{Comparison between the light curves for observing angle $\theta=20^\circ$
(red and magenta) and $\theta=25^\circ$ (green and blue).}
\end{figure}

\vfil\eject
\begin{figure}
\plotone{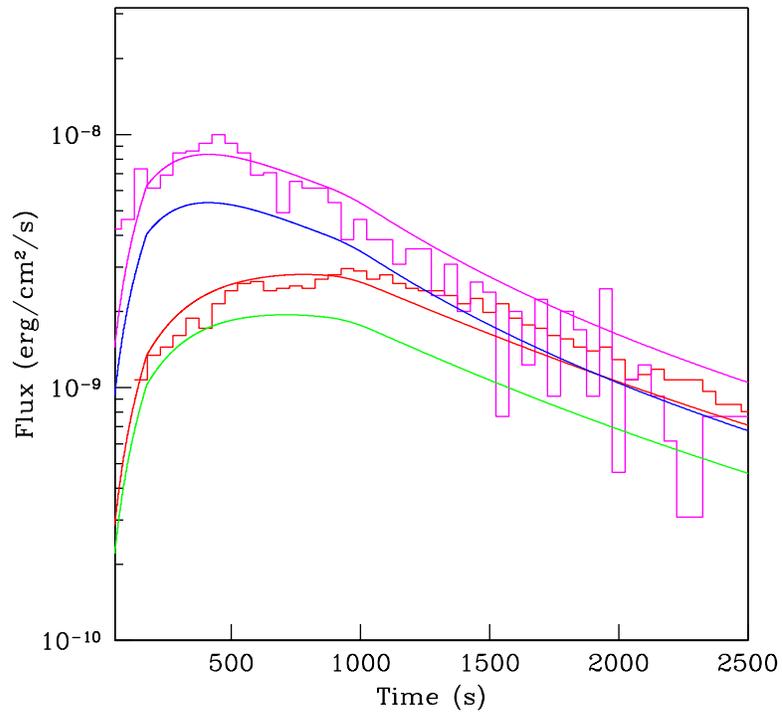} \caption{The light curves from a broken power law
 primary spectrum with different values of low energy index $\alpha=-0.9$
(red and magenta) and $\alpha=-1.1$ (green and blue).}
\end{figure}

\vfil\eject
\begin{figure}
\plotone{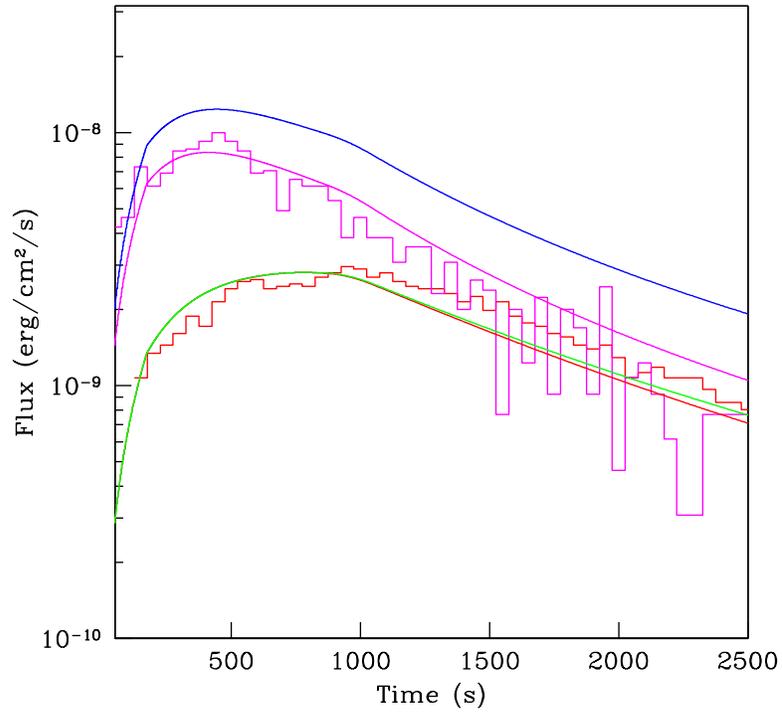}
\caption{The light curves from a broken power law primary spectrum
with different values of high energy index $\beta=-2.2$ (red and magenta)
and $\beta=-2.0$ (green and blue).}
\end{figure}

\end{document}